\documentclass[lettersize,journal]{IEEEtran}
\usepackage{amsmath,amsfonts}
 \usepackage[]{algorithm2e}

\usepackage{array}
\usepackage[caption=false,font=normalsize,labelfont=sf,textfont=sf]{subfig}
\usepackage{textcomp}
\usepackage{stfloats}
\usepackage{url}
\usepackage{verbatim}
\usepackage{graphicx}
\usepackage{cite}
\usepackage{xcolor}
\usepackage{lipsum}
\usepackage{hyperref}

\begin{document}

\title{Mitigating xApp conflicts for efficient network slicing in 6G O-RAN: a graph convolutional-based attention network approach}

\author{Sihem Bakri \IEEEauthorrefmark{1}, Indrakshi Dey \IEEEauthorrefmark{4}, Harun Siljak \IEEEauthorrefmark{1}, Marco Ruffini \IEEEauthorrefmark{1}, Nicola Marchetti \IEEEauthorrefmark{1}

\IEEEauthorblockA{\IEEEauthorrefmark{1} Trinity College Dublin, Dublin, Ireland;}
\IEEEauthorblockA{\IEEEauthorrefmark{4} Walton Institute of Information and Communication Science, Waterford, Ireland.}
\IEEEauthorblockA{Email: \IEEEauthorrefmark{1}sbakri@tcd.ie, \IEEEauthorrefmark{4} indrakshi.dey@waltoninstitute.ie, \IEEEauthorrefmark{1}harun.siljak@tcd.ie, \IEEEauthorrefmark{1}marco.ruffini@tcd.ie,  \IEEEauthorrefmark{1}nicola.marchetti@tcd.ie 
}

}

\maketitle

\begin{abstract}

O-RAN (Open-Radio Access Network) offers a flexible, open architecture for next-generation wireless networks. Network slicing within O-RAN allows network operators to create customized virtual networks, each tailored to meet the specific needs of a particular application or service. Efficiently managing these slices is crucial for future 6G networks. O-RAN introduces specialized software applications called xApps that manage different network functions. 
In network slicing, an xApp can be responsible for managing a separate network slice.
To optimize resource allocation across numerous network slices, these xApps  must coordinate. Traditional methods where all xApps communicate freely can lead to excessive overhead, hindering network performance. 
In this paper, we address the issue of xApp conflict mitigation by proposing an innovative Zero-Touch Management (ZTM) solution for radio resource management in O-RAN. Our approach leverages Multi-Agent Reinforcement Learning (MARL) to enable xApps to learn and optimize resource allocation without the need for constant manual intervention.  We introduce a Graph Convolutional Network (GCN)-based attention mechanism to streamline communication among xApps, reducing overhead and improving overall system efficiency. Our results compare traditional MARL, where all xApps communicate, against our MARL GCN-based attention method.  The findings demonstrate the superiority of our approach, especially as the number of xApps increases, ultimately providing a scalable and efficient solution for optimal network slicing management in O-RAN.


\end{abstract}


\section{Introduction}  

\IEEEPARstart{F}uture wireless networks such as Beyond 5G (B5G) and 6G are expected to provide significantly higher data rate, greater capacity, lower latency, higher coverage rate, and higher reliability than 5G networks.
They are expected to support the growing demand for high-speed, reliable, and secure communication services, particularly in increasingly data-intensive applications such as the Internet of Things (IoT), autonomous vehicles, and immersive virtual reality.
B5G/6G networks are also expected to offer revolutionary advances in wireless communications that will enable novel use cases and applications, such as remote surgery, holographic communication, and advanced manufacturing processes \cite{6GSurvey}.
Therefore, networks will need to be highly flexible, scalable, and adaptive to meet the dynamic and evolving demands of different applications and services. This will require a high degree of automation, virtualization, and intelligent network management \cite{6GSurvey}.

B5G/6G networks leverage Multi-Agent Reinforcement Learning (MARL) to train AI-powered software agents to collaborate and optimize network performance autonomously. This fosters adaptability, a critical quality when paired with ZTM, which empowers the network to self-configure based on predefined parameters, eliminating the need for constant human intervention. These features are particularly advantageous for Open-Radio Access Network (O-RAN) architectures, which promote vendor flexibility through software-defined solutions and the RAN Intelligent Controller (RIC) \cite{SON}.

O-RAN's inherent flexibility makes it an ideal foundation for network slicing, a key B5G/6G feature that enables the creation of virtual networks tailored to the specific needs of different applications and services. Each network slice can possess dedicated resources, bandwidth, and Quality of Service (QoS) to meet the demands of diverse applications, such as mission-critical industrial automation or ultra-reliable low-latency communications \cite{NS_in_6G}. However, managing a multitude of heterogeneous slices across a vast B5G/6G network presents a significant challenge. Network operators will require advanced automation techniques, intelligent solutions, and potentially even more sophisticated AI-powered applications to effectively handle the complexities of network slicing in modern communication systems \cite {O-RAN_Slicing_standard}.

In this paper, we propose a novel framework designed to provide flexible and autonomous network control, for capacity sharing across multiple xApps. The use of the RIC enables network slices to be managed by different xApps, which might be tuned to deliver different key performance targets. However, while this architecture offers flexibility in terms of independence of xApps design, it also creates potential conflict between xApps, as they share the same physical resources. In this work we address the issue of xApp conflict mitigation by developing ZTM of radio resource allocation between xApps that manage different network slices in an O-RAN scenario. Based on real-time analysis, the proposed system autonomously adjusts resource allocation among slices, according to the dynamic network condition and slice requirements.

 Our approach leverages Distributed Multi-Agent Reinforcement Learning, relying on Graph Convolutional Networks (GCN)-based attention network, as a key enabler for ZTM. The framework envisions a network where intelligence is embedded in the relationships between nodes, forming a complex and adaptive ecosystem. This paradigm shift moves away from traditional models, where intelligence resides within individual nodes, towards a more distributed and resilient approach.



The rest of this paper is organized as follows. Section II introduces the O-RAN architecture, including its main components, and also details the concept of Network Slicing in the O-RAN. Section III provides an overview of the main algorithms and approaches used to achieve a ZTM  of network slicing in the O-RAN architecture. Our proposed system model of conflict-aware radio resource management in O-RAN network slicing is then presented in Section IV. Following that, Section V presents the use case scenario, results, and analysis. Finally, we conclude this work in Section VI.

\section{Intelligent Network Slicing in O-RAN}
\label{sec:ORAN}

This section describes how the overall O-RAN architecture, characterised by the use of the RAN Intelligent Controller, enables ZTM of dynamic network slicing.


\subsection{O-RAN Architecture}

\begin{figure}
    \includegraphics[width=1.6\columnwidth]{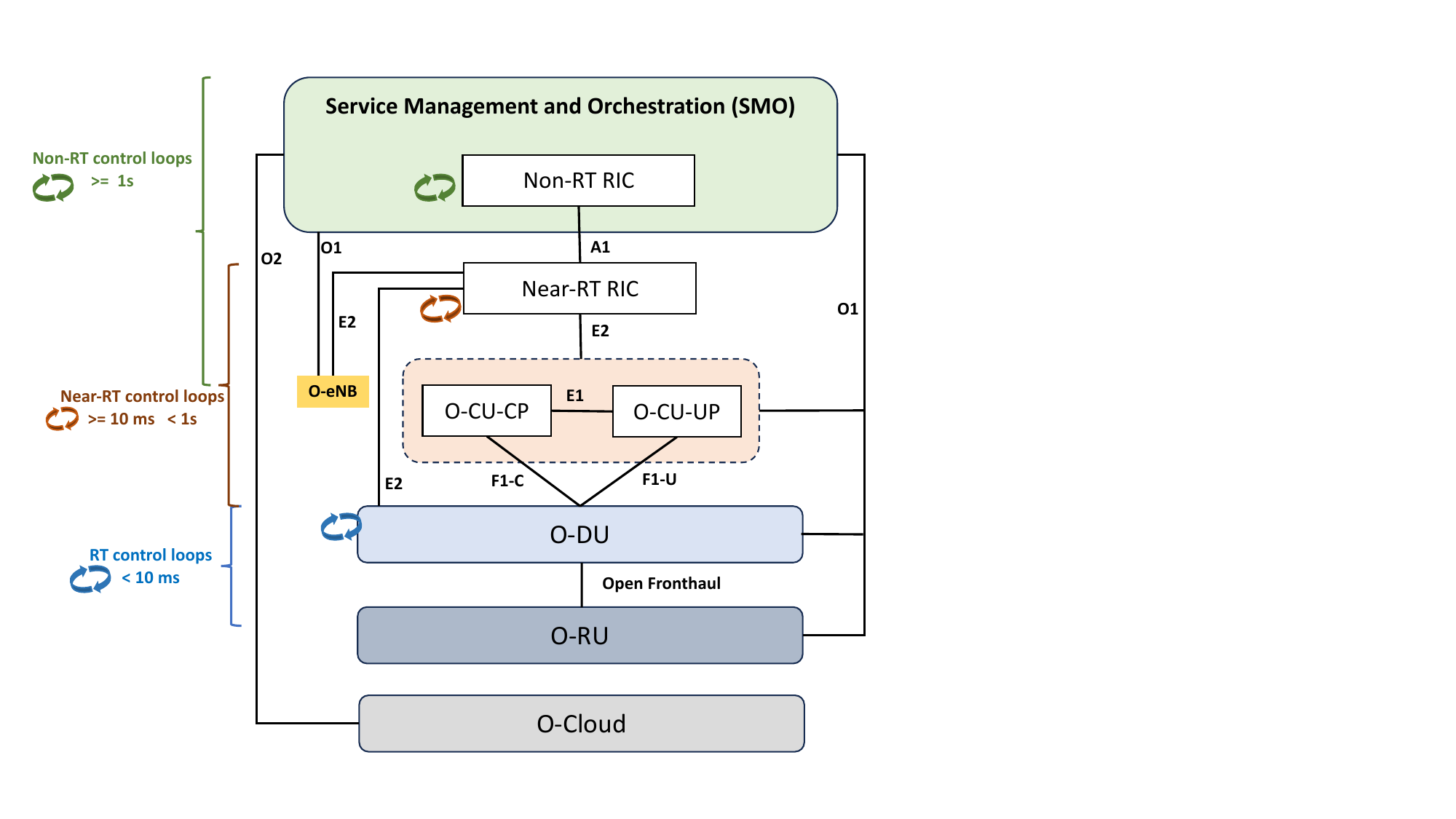}
    \caption{An Overview of O-RAN Architecture: Components, Control Loops and Interfaces.}
    \label{fig:ORAN}
\end{figure}

O-RAN is transforming the telecommunications industry by disaggregating traditional RAN components. This allows the use of open, interoperable hardware and software from diverse vendors, fostering flexibility, scalability, and innovation. O-RAN aims to improve performance, reduce costs, and streamline network management. O-RAN employs closed-loop control functionalities that dynamically optimize various aspects of the network. These include Non-Real Time (Non-RT), Near-Real Time (Near-RT), and Real-Time (RT) control loops as depicted in Fig. \ref{fig:ORAN}, each coordinating different parts of the system to enhance flexibility and interoperability \cite{O-RAN_standard}. Key components within O-RAN architecture comprises of:
\begin{itemize}
    \item \emph{\textbf{Service Management and Orchestration (SMO)}}: Manages network infrastructure, providing interfaces for monitoring and control. The SMO supports network functions, optimization, and workflow management.
    \item \emph{\textbf{Radio Intelligence Controller (RIC)}}: Enables intelligent applications such as intelligent traffic steering. It gathers data from O-RAN units, allowing control apps (xApps and rApps) to analyze information and optimize radio resources for improved network performance. The RIC comprises of two key components: a) \emph{Near-Real Time RIC (Near-RT RIC) -} Operates on control loops between 10ms and 1s for rapid network adjustments. b) \emph{Non-Real Time RIC (Non-RT RIC) -} Operates on control loops above 1s for broader optimization strategies.
\end{itemize}

The O-RAN architecture is characterized by openness and flexibility, enabled by multiple \textbf{\textit{inter-working functional entities}}: O-RU, O-DU, O-CU, O-eNB and O-cloud. The O-RU represents the physical layer, incorporating remote radio units for transmission and reception. The O-DU serves as a distributed processing unit, managing baseband processing for multiple O-RUs. O-CU is the centralized unit responsible for higher-layer processing, orchestrating resources across multiple O-DUs. O-eNB is the evolution of the traditional base station, adapting to the open and interoperable principles of O-RAN. O-Cloud signifies the cloud infrastructure in the O-RAN ecosystem, providing a virtualized environment for network functions. These entities jointly enable the agility, interoperability, and evolution in O-RAN.

 Another fundamental aspect of O-RAN architecture is functional split aimed at disaggregating RAN components to enhance interoperability among equipment from various vendors.  Different functional split options between CU and DU have been studied and specified, ranging from higher-layer to lower-layer splits. Split 7.2 is one of the commonly adopted options, defining the distribution of processing tasks between CU and DU \cite{O-RAN_wg4}.

The O-RAN functions communicate between them through standardised interfaces including \textbf{\textit{A1, O1, O2, E2, Y1 , O-Cloud Notification}}, and \textbf{\textit{Open Fronthaul interfaces}} as shown in Fig. \ref{fig:ORAN}.

%

\subsection{Network Slicing in O-RAN}

Network slicing enables the creation of multiple, independent virtual networks on a shared physical infrastructure. This technique offers benefits like multi-tenancy, allowing different virtual networks to share resources and reduce costs. It also provides service isolation, guaranteeing performance for diverse services with specific Service Level Agreements (SLAs). Moreover, network slicing enhances flexibility by enabling on-demand creation, modification, or deletion of slices, making the network highly adaptable to changing needs and requirements \cite{O-RAN_Slicing_standard} \cite{NS}.

Network slicing enables end-to-end control of network resources, including core, transport, and RAN. While 3GPP started defining network slicing support with Release 15, further O-RAN architecture modifications are required to fully implement slicing in an open RAN environment. 
The O-RAN reference slicing architecture incorporates slice management functions along with O-RAN architectural components.
The key functions that contribute to the set of network slicing in O-RAN are: Non-RT RIC, Near-RT RIC, O-CU, O-DU, and A1, A2, O1, E2 interfaces; where their role and impact are described below. Furthermore, we present the transport layer operations required to implement network slicing in O-RAN.

\subsubsection{Non-RT RIC } 

The non-RT RIC, a crucial component in the O-RAN slicing architecture, serves as the gateway to gather important data for the SMO framework. Its interaction with third party algorithms, often based on AI/ML, helps define innovative RAN slicing use cases. The non-RT RIC retrieves information such as subnet SLAs, performance indicators, and other attributes from SMO, to construct AI/ML models for deployment in Near-RT RIC. It uses this information to optimize slice parameters for Near-RT RIC, O-CU, and O-DU via O1 interface interaction with SMO. 

\subsubsection{Near-RT RIC} 
In the O-RAN slicing architecture, the Near-RT RIC plays a vital role in near-real-time RAN slice subnet optimization. It utilizes xApps that communicate parameters to the O-CU and O-DU through the E2 interface, leveraging AI/ML models or other control mechanisms. The Near-RT RIC receives slice subnet information during provisioning via the O1 interface (similar to the Non-RT RIC), aiding in the support of sliced RAN resources. Slice resource configuration on E2 nodes occurs through O1 configuration with the SMO and via E2 configuration in two loops. SLA assurance parameters from the A1 interface guide the Near-RT RIC in ensuring optimal E2 node behavior. This component also handles fast-loop configuration and O1 slice-related configuration (like RRM policy data to the O-CU), all configured via the SMO. To achieve this, the Near-RT RIC relies on slice-specific, near-real-time performance data from the E2 interface, necessitating robust Performance Management (PM) mechanisms between E2 nodes and the Near-RT RIC.

\subsubsection{O-CU} 
The O-CU consists of a single control plane entity, the O-RAN CU–Control Plane (O-CU-CP) and potentially multiple user plane entities, the O-RAN CU–User Plane (O-CU-UP), communicating via the E1 interface. It is essential for the O-CU to support slicing features outlined by 3GPP such as mobility management.

\subsubsection{O-DU} 
The O-DU, responsible for running lower-layer RAN protocols, is tasked with supporting slice-specific resource allocation strategies. In alignment with initial O1 configuration for Physical Resource Block (PRB) allocation levels, coupled with directives from O-CU over the F1 interface, and dynamic guidance from Near-RT RIC via the E2 interface, the MAC layer must allocate and isolate pertinent PRBs for specific slices.

\subsubsection{A1 Interface}
 The A1 interface, connecting the Non-RT RIC and the Near-RT RIC, supports policy management, ML model management, and enrichment information services, and is essential for various slicing use cases like slice SLA assurance. The non-RT RIC employs policy management to send slice-specific policies guiding the Near-RT RIC in resource allocation and control activities, and to receive policy feedback.

\subsubsection{E2 Interface}

E2, serving as the interface between the Near-RT RIC and E2 nodes, facilitates E2 primitives (Report, Insert, Control, and Policy) to govern services provided by E2 nodes such as Radio Resource Management service. These primitives will be employed by slice-specific applications (xApps) to influence E2 nodes' slice configurations, addressing slice-centric radio resource management, resource allocations, MAC scheduling policies, and configuration parameters within RAN protocol stacks.

\subsubsection{O1 Interface}

O1, serving as the interface between O-RAN managed elements and the management entity, adheres to specifications in \cite{13_MIS}. It configures O-RAN nodes' slice-specific parameters based on the slice's service requirements, utilizing information models outlined in \cite{7_3GPP}. These models include RRM policy attributes, such as PRB ratio and PRB split among slices. To accommodate O-RAN slicing use cases, 3GPP information models may be extended or defined to capture slice profiles and specific configuration parameters, transmitted over the O1 interface.
Additionally, O1 is employed to configure and collect slice-specific performance metrics such as throughput and latency, and faults such as RAN failure in O-RAN nodes.

\subsubsection{O2 Interface}

O2, the interface connecting SMO and O-Cloud, focuses on the lifecycle management of virtual O-RAN network functions. During the creation and provisioning of RAN Network Slice Subnet Instances (NSSI), the RAN Network Slice Subnet Management Function (NSSMF) collaborates with SMO to initiate the instantiation of essential O-RAN functions (such as Near-RT RIC, O-CU-CP, O-CU-UP, and O-DU) aligned with slice requirements. 
After RAN NSSI creation and provisioning, NSSMF can execute NSSI modification and deletion procedures through interaction with SMO.
However, O2 is not anticipated to handle the lifecycle management of Non-RT RIC, given its integral association with SMO.

\subsubsection{ Transport Network Slicing} 
This involves addressing the Fronthaul (FH) interface between O-RU and O-DU, and the Midhaul (MH) interface between O-DU and O-CU. Various approaches are emerging to define transport network slicing in alignment with 5G requirements. 
These approaches address how mobile interfaces (Fronthaul and Midhaul within RAN slice subnet, Backhaul between RAN slice subnet and Core slice subnet, and between Core slice subnet and the Packet Data Network (PDN)) can be sliced, determining their structure, and specifying the number of slices needed at the transport layer.

\section{ZTM Based GCN-MARL}

\label{sec:Background}

\subsection{Multi Agent Reinforcement Learning }

A MARL system describes multiple distributed entities—so-called agents—which take decisions autonomously, and interact within a shared environment through reinforcement learning. 
Unlike single-agent reinforcement learning, where an individual agent learns to make decisions based on its own actions and rewards, MARL employs multiple autonomous agents that simultaneously learn and adjust their behaviors by taking into account the actions and rewards of other agents.

In this context the agents behaviors could be fully cooperative, fully competitive and a combination of the two, depending on the interaction dynamics of the agents. In cooperative setting, agents collaborate to achieve a common goal by sharing information and coordinating their actions to maximize collective rewards. The competitive MARL involves agents with conflicting goals competing for limited resources or rewards, and aiming to outperform others through learned strategies. 
The above settings i.e. cooperative and competitive, introduce complex interactions and dependencies, which makes MARL well adapted to real-world scenarios for instance: multi-robot systems, autonomous vehicles, networked communication, and strategic games, among others \cite{MARL} \cite{MARL2}.

MARL algorithms fall into two main categories: centralized, where a single controller coordinates actions for all agents (simple coordination, poor scalability), and decentralized, where agents act based on their own observations (scalable, but complex coordination). A hybrid approach, centralized training and distributed execution, seeks to leverage the best of both worlds. During training, agents share information globally, improving cooperation and learning. In execution, they act independently based on learned policies, allowing for adaptability and scalability \cite{MARL2}. 

To address the challenges of MARL, various algorithms have been proposed such us Q-learning, policy gradient, and game theoretic concepts including Nash equilibrium and Markov games, that have been applied to model and analyze the interactions between agents in MARL.

\subsection{Graph Convolutional Network}

A GCN is a type of neural network designed to process data on graphs, where each node is represented by a high-dimensional feature vector,  whose dimension depends on the input data, specifically the number of attributes (features) associated with each node.
GCNs leverage the local structure of a graph by using adjacency matrices or other connectivity-based representations to define node relationships \cite{GCN}.
Typically, the adjacency matrix is normalized to improve stability during training and prevent vanishing or exploding gradients.
The convolutional layers in GCN are designed to capture the neighborhood information of each node.
In fact, each layer aggregates information from the immediate neighborhood of each node and updates the node's representation, namely a high-dimensional feature vector.
Unlike traditional CNNs, GCNs use a message-passing mechanism rather than spatial convolutions, where each node iteratively aggregates features from its neighbors.
The information aggregation process takes into account the node's features as well as the features of its neighboring nodes, allowing GCNs to capture complex relationships and dependencies in the graph.
When multiple GCN layers are stacked, nodes can incorporate information from multi-hop neighbors, enabling a more global understanding of the graph structure \cite{GCN}.

\subsection{Attention networks}


Attention mechanisms correspond to a layer within Neural Networks (NN) designed to selectively focus on specific parts of data, enhancing the network's ability to prioritize relevant information.
This NN layer transforms the input features into a series of output features, which are the attention weights. 
These weights determine the importance of different input features at each step in the process, as shown in Fig. \ref{fig:GAN}.

There are several types of attention networks, such as Transformers, Multi-head attention, and Self-attention. In this study we applied the Multi-Head Attention (MHA) network, since it allows the model to focus on different parts of the input sequence simultaneously, with multiple attention heads computing attention weights independently \cite{Attention}. The MHA process is described in \cite{Attention}.

\begin{figure*}
    \includegraphics[width=0.98\textwidth]{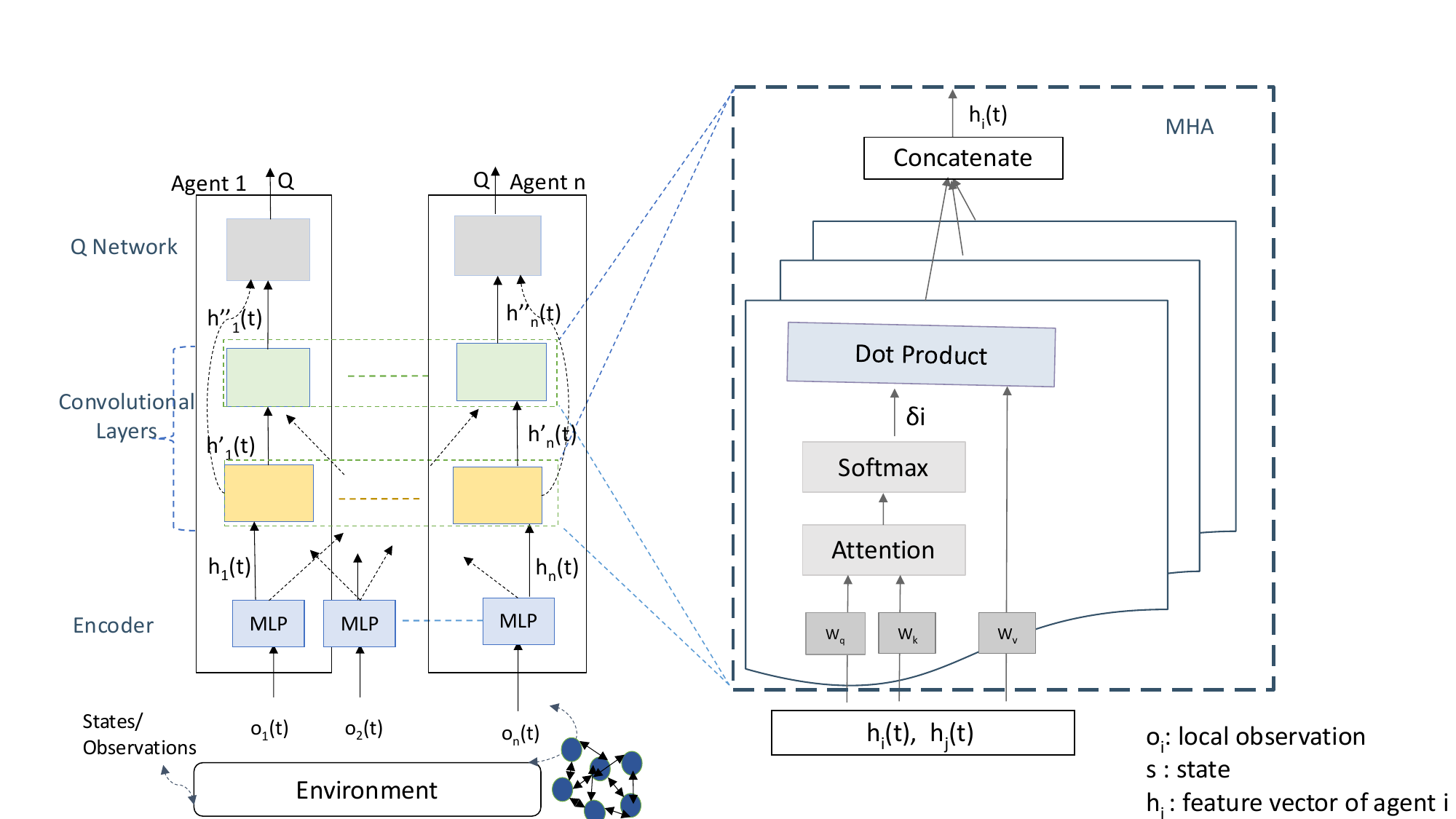}
    \label{fig:GAN}
 \caption{
Graph Convolutional Network-Based Attention Modules: in this illustration, each node i.e., xApp collects observations, which are then encoded by an MLP. The output of the MLP is then processed by two successive convolutional layers incorporating multi-head attention (MHA). Finally, the resulting output of these convolutional layers serves as input to the Q network. }
    \label{fig:GAN}
\end{figure*}


\section{Conflict-Aware Radio Resource Management in O-RAN: System Model}



\label{sec:GCN-MARL}

In this paper, we propose a conflict-aware resource management framework to mitigate xApp conflict while optimizing radio resource allocation within the O-RAN architecture.
In fact, in O-RAN, an xApp can manage a network slice. Therefore, xApps that manage network slices must communicate with each other. However, key challenges arise as the number of xApps increases, including high communication overhead and potential interference in xApp interactions. To address this challenge, we adopt a MARL algorithm to model our architecture, where each xApp manages a slice and is controlled by an agent.

To alleviate the above-mentioned concern, we develop a model that captures environmental dynamics and reduces communication overhead between agents while maintaining high communication efficiency. To this end, we propose applying a GCN enhanced with MHA to strengthen the MARL framework. Such models can also capture how nodes and links respond to changes in communication scenarios, network conditions, and environmental variations.
GCN-Based Attention—primarily MHA— is a hybrid approach that combines the strengths of graph-based convolutional operations and attention mechanisms in GNNs. This method incorporates attention mechanisms into the message-passing process of GCN, as shown in Fig. \ref{fig:GAN}, allowing nodes to aggregate features not only from neighboring nodes but also to focus on specific neighbors using attention weights. By integrating attention into the convolutional process, the model learns more expressive and adaptive representations, capturing both local and global graph structure information. In this case, agents only communicate with the most relevant neighbors instead of all agents in the network. As a result, multi-agent systems reduce overhead caused by unnecessary communication between all agents.
Each slice is controlled by an agent within a multi-agent system, with the number of agents scaling dynamically with the number of slices. The GCN-based attention network addresses communication challenges in three key steps: (i) State Encoder: each agent pre-processes its observations using a Multi-Layer Perceptron (MLP) to extract feature vectors; (ii) Convolutional Attention Network: agents’ feature vectors are merged using adjacency matrices and convolutional layers with MHA to enhance cooperation, extract latent features, and capture relationships between agents; and (iii) Q-Network: Latent feature vectors are aggregated to compute action weights, enabling collaboration between distant agents and improving decision-making by leveraging diverse information sources, such as local network conditions, agent states, and neighboring agents’ states, for efficient PRB sharing. 
We adopt Deep Q-Learning to train our model, where future value estimation serves as the target for the current estimation, ensuring stable cooperation even as the states of surrounding agents change. In this context, the attention weight distribution in the next state is used as the target for the current attention weight distribution \cite{DGN}.

\section{Use case scenario and Analysis}

\subsection{Scenario Description}

Our system initially models the network environment as a graph, where each node is represented by an agent. The considered state, actions and reward for our MARL algorithm are designed as follows: 

\subsubsection {States or Observations} consist of the current available resources in the infrastructure provider, along with a range of information for each slice. This includes the current state of the slice channel, mainly the Channel Quality Indicator (CQI), Modulation and Coding Scheme (MCS), and the throughput requirements of each slice. Any other relevant metrics that can help estimate the necessary number of PRBs for each slice could be included here.


\subsubsection {Actions} the number of resources to assign to each network slice.


\subsubsection {Reward function} depends on the slice's required SLA specifications in terms of throughput and slice priority. Additionally, it penalizes the allocation of excessive resources to a slice when it does not require them. The reward function is the following:
$ Reward= min (\frac{T_{req}}{T_{alloc}},1) \cdot P_{slice} - Penalty$; 

where $T_{req}$ and $T_{alloc}$ are the required and allocated throughput respectively and $T_{alloc}$ $\neq$ 0 ; $P_{slice}$ represents the priority level assigned to each slice, with values set to 1, 2, or 3, indicating low, medium, and high priority, respectively. The $Penalty$ term equal to $ max (0, ({T_{alloc}}-{T_{req}}))$, it penalizes the agent for allocating excessive resources to a slice, i.e., beyond its actual requirements.


\vspace{5mm}

Our algorithm operates within the O-RAN architecture, specifically at the RIC level. We assume that each node, i.e., a slice, is managed by an agent representing an xApp in the Near-RT RIC. The training process is supposed to run at SMO level, within the non-RT RIC entity.
In O-RAN, xApps in the Near-Real-Time RIC  communicate through well-defined interfaces and protocols as described in Section \ref{sec:ORAN}. These interfaces enable information exchange and coordination among different xApps, to collectively optimize and control the radio access network. The E2 interface handles interactions between RIC and gNB, while the A1 interface facilitates policy-based communication. These standardized interfaces ensure interoperability and seamless collaboration among xApps within the MARL architecture.

In our approach, the PRBs requirements of each activated slice vary dynamically and remain unknown to the xApps until they receive updated information from the respective slice. The xApps may control parameters such as the required number of PRBs over time, the number of users in each slice, and other slice information like channel availability, security, and access control. 
Therefore, at each time step $t$, each slice agent $a$ at xApp $i$ receives local observations $o^{i}(t)$ as described above. 
The observations obtained are then pre-processed, and the xApps exchange information following our proposed process described in Section \ref{sec:GCN-MARL}.
Our proposed scheme allows to avoid high interference between xApps, since only some agents communicate with each other.


\subsection{Experimental results}

To test our approach we used dataset shared by the COLOSSEUM platform \cite{colosseum}.
 COLOSSEUM  is the world's largest wireless network emulator, designed to facilitate the exploration and advancement of large-scale, next-generation radio network technologies in repeatable and highly configurable Radio Frequency (RF) and traffic environments for research and development purposes. The dataset used in our study comes mainly from the following network configuration, which includes 7 Base Stations (BSs). The channel bandwidth is 10 MHz (50 PRBs). Each BS has 3 slices: eMBB, mMTC and URLLC. 
The traffic classes of each slice are Constant Bit-Rate traffic (4 Mbps per UE) for eMBB, Poisson traffic (30 packets/s of 125 bytes per UE) for mMTC, and Poisson traffic (10 packets/s of 125 bytes per UE) for URLLC.
The UEs are static and uniformly distributed within 50 m of each BS.

To create a more complex environment, we assume that our environment is composed of 10 slices: 4 eMBBs, 3 mMTCs, and 3 URLLCs. 
We assume that the channel quality of a User Equipment (UE) is either similar or identical to the channel quality of the slice to which it belongs.

\begin{figure}
    \centering
    \includegraphics[width=\columnwidth]{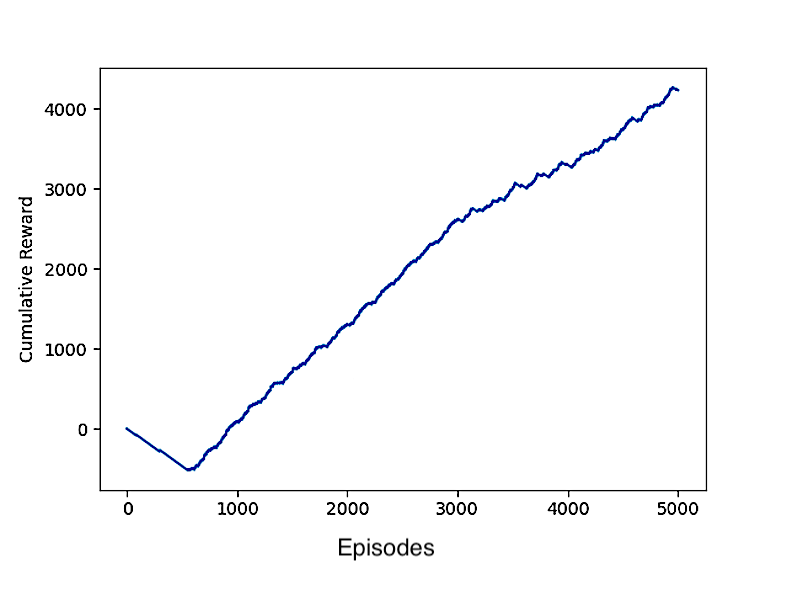}
    \caption{The cumulative reward obtained during the training of MARL-GCN-based attention over 5000 episodes.}
    \label{fig:reward}
\end{figure}

In Fig. \ref{fig:reward} we simulate the cumulative reward, to observe how agents act over time to achieve their goals during the training phase.
Initially, we observe a decrease in reward, followed by an increase. This pattern indicates a dynamic learning process in which agents first explore sub-optimal strategies before adapting and converging on a more efficient one, thereby maximizing rewards.

\begin{figure}
    \centering

    \includegraphics[scale=0.49]{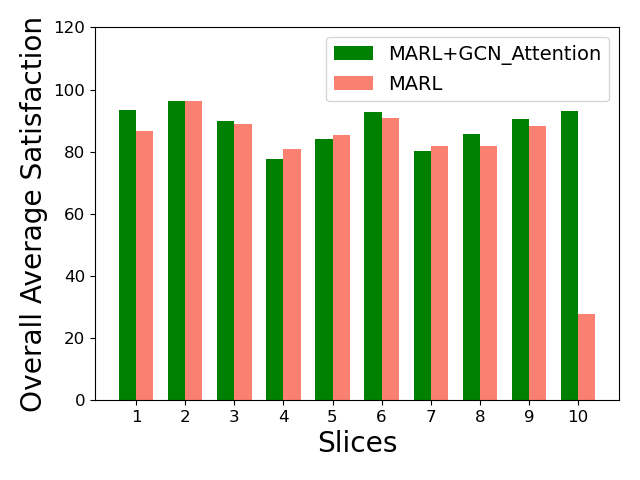}
    
    \caption{The satisfaction factor of 10 activated slices using MARL-based GCN and the traditional cooperative MARL (the ratio of the number of PRBs allocated to the requested according to throughput).}
    \label{fig:comparison}
\end{figure}

In Fig. \ref{fig:comparison}, we simulate the average satisfaction of each slice over time in terms of throughput using our proposed approach, namely MARL-based GCN, compared to the cooperative MARL algorithm. The satisfaction factor represents the ratio of the number of allocated PRBs to the requested PRBs based on throughput.
The obtained histogram shows that the standard cooperative MARL algorithm—where all agents communicate between them—achieves good satisfaction levels, with 9 slices maintaining satisfaction above 80\% and one slice around 30\%. This indicates that while the method generally performs well, it may lead to imbalances in resource allocation. Employing GCN-based attention networks within MARL improves performance further. In this case, 9 slices maintain satisfaction above 80\%, while the remaining slice reaches approximately 75\%, despite only neighboring agents communicating. This demonstrates that structured, selective communication in MARL—enabled by attention mechanisms—can enhance fairness and efficiency in resource allocation.

\begin{figure}
    \centering
    \includegraphics[scale=0.49]{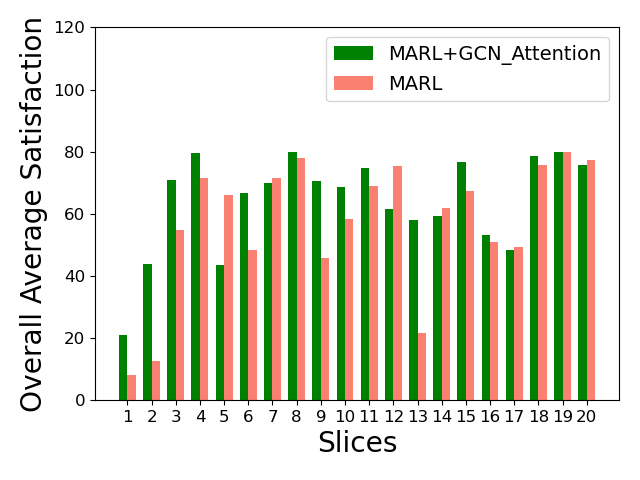}
    \caption{The satisfaction factor of 20 activated slices using MARL-based GCN and the traditional cooperative MARL (the ratio of the number of PRBs allocated to the requested according to throughput).}
    \label{fig:comparison_2}
\end{figure}

In Fig. \ref{fig:comparison_2}, we simulate the same scenario as depicted in Fig. \ref{fig:comparison}, but with 20 slices instead of 10. We observe that with 20 slices—i.e., 20 xApps communicating with each other—the MARL algorithm with GCN outperforms traditional MARL. This improvement is attributed to GCN-based attention, which assigns weights to slices based on their importance (i.e., the number of PRBs required), rather than treating all slices equally, as in traditional MARL. By prioritizing resource allocation dynamically, our approach leads to better overall network performance. Additionally, we note that satisfaction levels decrease when managing 20 slices compared to 10, which is expected due to bandwidth limitations in terms of available PRBs. As the number of slices increases, the finite bandwidth may not be sufficient to fully satisfy all demands, leading to a reduction in overall satisfaction.

While traditional cooperative MARL models achieve good results, they may suffer from communication conflicts and overhead as the number of xApps increases. In contrast, incorporating GCN-based attention in MARL improves performance, even with a higher number of agents. This is primarily due to the ability of GCN-based attention to assign dynamic weights to xApps based on their features and structural relationships, allowing the network to prioritize more critical slices.

From Fig. \ref{fig:comparison} and Fig. \ref{fig:comparison_2}, we conclude that MARL-GCN-based attention handles large-scale slicing environments more efficiently than traditional cooperative MARL models, which rely on explicit communication. This efficiency results from MARL-GCN’s capability to propagate relevant information across the xApp graph while significantly reducing communication overhead.


To evaluate the xApps communication overhead, we conduct the following analysis. In the MARL algorithm, communication between xApps (slices) occurs at a rate of $n (n-1)$ per timestep, where $n$ represents the number of slices. However, in MARL-GCN, each xApp communicates with only $k$ neighbors, where $k < (n-1)$, the total number of communications becomes $n \cdot k$. As a result, the overhead is reduced by $\frac{(n-1)-k}{(n-1)} \times 100$\%, as compared to the traditional approach.
\par For instance, in our 10-slice scenario using traditional MARL, there are 90 communications between xApps. Conversely, in MARL with GCN, for instance, if each agent exchanges information with only 3 other agents, the total number of communications goes down to 30, resulting in a 66\% reduction in overhead.
Based on the above, we can infer that employing MARL along with GCN, significantly reduces the overhead between xApps, consequently mitigating xApp interference.

\section{Conclusion}

This paper proposes a ZTM solution to address xApp communication conflicts that arise when managing radio resources in network slicing within the O-RAN architecture. The proposed approach leverages a MARL framework, where each agent represents an xApp responsible for managing a network slice. By integrating a GCN-based attention mechanism into MARL, our solution enhances dynamic resource allocation while mitigating xApp conflicts, particularly as the number of slices increases. This structured communication framework ensures that xApps interact selectively, reducing unnecessary overhead and improving coordination in resource distribution.

The obtained results demonstrate that GCN-based attention in MARL improves scalability and efficiency compared to traditional cooperative MARL, making it a promising approach for conflict-aware resource management in O-RAN network slicing.

In future work, we aim to extend the proposed solution to achieve ZTM across end-to-end network slices, encompassing the core, transport, and radio access networks within a real O-RAN architecture testbed in our lab.

\section*{Acknowledgement}

This work is supported by Science Foundation Ireland (SFI) and is co-funded under the European Regional Development Fund under Grant Numbers  13/RC/2077-P2.

The work of I. Dey is partly supported by EU ERDF Project "SAtComm" under Grant Number EAPA\_0019/2022 and EU MSCA Project "COALESCE" under Grant Number 101130739.

\vfill

\end{document}